\documentclass{physeauth}
\usepackage{graphicx}
\usepackage{amsmath}
\usepackage{amssymb}

\begin{document}

\begin{frontmatter}

\title{Impurity problems for steady-state nonequilibrium
dynamical mean-field theory}

\author{J. K. Freericks\thanksref{thank1}}

\address{Department of Physics, Georgetown University, Washington, DC 20057,
USA}

\thanks[thank1]{
Corresponding author. 
E-mail: freericks at physics dot georgetown dot edu}

\begin{abstract}
The mapping of steady-state nonequilibrium dynamical mean-field theory 
from the lattice to the impurity is described in detail.  Our
focus is on the case with current flow under a constant dc electric
field of arbitrary magnitude.  In addition to formulating the problem
via path integrals and functional derivatives, we also describe the
distribution function dependence of the retarded and advanced Green's
functions. Our formal developments are exact for the Falicov-Kimball
model.  We also show how these formal developments are modified for more
complicated models (like the Hubbard model).
\end{abstract}

\end{frontmatter}


\section{Introduction}

Dynamical mean-field theory (DMFT) began in 1989, when Metzner and Vollhardt suggested 
the large-dimensional limit (with an appropriate rescaling of the hopping
integral) as a simplifying limit for the many-body problem that nevertheless still included
the important competition between minimizing the kinetic and potential energies of the system~\cite{metzner_vollhardt}.
These ideas were implemented into a complete DMFT by Brandt and Mielsch with the exact solution of the Falicov-Kimball model at half filling~\cite{brandt_mielsch}.  Since that time, nearly all equilibrium many-body models have been solved in the large-dimensional limit with DMFT.  Recently, progress has been made on nonequilibrium extensions of DMFT~\cite{freericks_prl,freericks_prb,kollar_eckstein,tran,aoki} for the Falicov-Kimball model and for the
Hubbard model~\cite{freericks_hubbard_prl}.  In situations where the perturbing fields (or the time-dependent part of the Hamiltonian) maintains the translational invariance of the lattice, the self-energy remains uniform and local, as follows from the Langreth rules~\cite{langreth} and the perturbative expansion in the hopping by Metzner~\cite{metzner}.

In this work, we examine the steady-state limit of nonequilibrium DMFT.  We start the system in equilibrium at an inverse temperature $\beta$ and wait a long time after the field or time-dependence has been turned on, so that the system has reorganized itself to the long-time response of the driving fields (or time dependence)~\cite{kadanoff_baym,keldysh}.  We are inherently assuming that the long-time limit of the Hamiltonian is different from the original equilibrium Hamiltonian.  It need not have any explicit time dependence in this limit though.  We will show that the Keldysh formulation of the many-body theory \{ignoring the third (imaginary) branch of the contour described by Wagner~\cite{wagner}\} produces the exact steady-state solution for the Falicov-Kimball model; we will also discuss modifications for the Hubbard model (and more complicated models).  Our approach works with functional integrals and derivatives.  Most of the techniques are completely straightforward, but these details have not appeared in the literature yet, and provide useful insights into the solution of nonequilibrium problems via DMFT.

\section{Formalism for the Falicov-Kimball model}
\label{sec:fk}

We will focus our efforts first on the Falicov-Kimball model, whose Hamiltonian~\cite{falicov_kimball} is the following [in a uniform electric field described by a uniform vector potential ${\bf E}=-\partial {\bf A}(t)/c\partial t$]:
\begin{equation}
 \mathcal{H}_{FK}=\sum_{\bf k}\{\epsilon[{\bf k}-e{\bf A}(t)/\hbar c]-\mu\}c^\dagger_{\bf k}c^{}_{\bf k}+
U\sum_ic^\dagger_ic^{}_iw_i,
\label{eq: ham_fk}
\end{equation}
where we employed the Peierls' substitution~\cite{peierls}.  Here, $c^{}_{\bf k}$ and $c^{\dagger}_{\bf k}$ destroy and create a spinless itinerant fermion with momentum ${\bf k}$, $\epsilon({\bf k})=-\lim_{d\rightarrow\infty}(t^*/\sqrt{d})$ $\sum_{i=1}^d\cos {\bf k}_i$ is the bandstructure on an infinite-dimensional hypercubic lattice, $\mu$ is the itinerant-electron chemical potential, $U$ is the conduction-electron--local-electron interaction, $w_i$ is the number operator for the spinless localized electrons at site $i$, and we use the real-space basis for the itinerant electron operators in the second term of $\mathcal{H}$. In the following we set $\hbar=c=t^*=1$.

We are interested in the steady-state limit of the response, which can be determined in the limit where the vector potential is turned on in the infinite past (but after the system has fully equilibrated into an equilibrium distribution with inverse temperature $\beta$); in other words, we choose ${\bf A}(t)=\lim_{t_0\rightarrow-\infty}\theta(t-t_0)(t-t_0){\bf E}$. In nonequilibrium physics there are two different Green's functions that are required to describe the system.  One is the retarded Green's function and the other is the Keldysh Green's function.  The former describes how the quantum density of states (DOS) varies with energy, while the latter describes how the electrons are distributed amongst those states.  Naively, one would assume that the quantum states would be independent of the initial temperature, although they will depend on the electric field strength, the fillings of the particles, and the interaction energy.  But it is well known in equilibrium many-body physics that interacting systems often display some temperature dependence to the DOS. Noninteracting problems, and the Falicov-Kimball model are exceptions, however, where the DOS is independent of how the electrons are distributed amongst the states.  It is the goal of this work to elaborate on this situation in nonequilibrium cases.

Our starting point is to define the contour-ordered Green's function on the two-branch Keldysh contour, which runs from
$t=-\infty$ to $t=\infty$ and back.  We define it only for the case of an impurity in a time-dependent dynamical mean field denoted $\lambda_C$:
\begin{equation}
G_C(t,t^\prime)=-i{\rm Tr}_{cf} \{\mathcal{T}_C \rho_{imp}S(\lambda_C)c(t)c^\dagger(t^\prime)\}/\mathcal{Z},
\label{eq: g_contour_def}
\end{equation}
with $\rho_{imp}$ the steady-state density matrix for the impurity, the operators denote time evolution in the Heisenberg picture with respect to the impurity Hamiltonian $\mathcal{H}_{imp}=-\mu c^\dagger c +Uw_ic^\dagger c$,
the S-matrix (evolution operator) satisfies
\begin{equation}
 S(\lambda_C)=\mathcal{T}_C\exp\left [ -i\int_Cd\bar t\int_Cd\bar t^\prime c^\dagger(\bar t)\lambda_C(\bar t,\bar t^\prime)c(\bar t^\prime)\right ],
\label{eq: sdef}
\end{equation}
$\mathcal{Z}={\rm Tr}_{cf}\rho_{imp}\mathcal{T}_CS(\lambda_C)$ is the partition function, the time ordering is with respect to the ordering
along the contour, and the trace is over the four states of the spinless conduction and localized electrons. The long-time-limit density matrix is {\it a priori} unknown (while the lambda field is determined self-consistently with the DMFT algorithm).  In practice, one needs to evolve the full many-body system from the time the field is turned on until the long-time limit is reached (and use the so-called restart theorem~\cite{restart} to determine it).  We will see, however, that the density matrix does not enter into the calculation of retarded or advanced quantities for the Falicov-Kimball model, so we defer further discussion at the moment. We will work with Wigner coordinates~\cite{wigner} of average $T=(t+t^\prime)/2$ and relative $t_{rel}=t-t^\prime$ time below.

We begin our discussion by considering a problem that has no localized particles, so we do not take the trace over the $f$-electrons, but instead, simply set $w_i=0$.  The full solution, including the trace over $f$-particles can be easily constructed from this solution, as we show below.  It is convenient to break up the two-branch contour into a $+$ branch, where the time increases from $-\infty$ to $+\infty$ and a minus branch, where time decreases from $+\infty$ to $-\infty$. Then the evolution operator in Eq.~(\ref{eq: sdef}) can be rewritten as
\begin{eqnarray}
S(\lambda_C)&=& \mathcal{T}_C\exp \Biggr [ -i\int_{-\infty}^\infty d\bar t\int_{-\infty}^\infty d\bar t^\prime
\Biggr \{ \label{eq: s_tlgtbar}\\
&~&c_+^\dagger(\bar t)\lambda^T(\bar t,\bar t^\prime)c^{}_+(\bar t^\prime)
-c_+^\dagger(\bar t)\lambda^<(\bar t,\bar t^\prime)c^{}_-(\bar t^\prime)\nonumber\\
&-&c_-^\dagger(\bar t)\lambda^>(\bar t,\bar t^\prime)c^{}_+(\bar t^\prime)
+c_-^\dagger(\bar t)\lambda^{\bar T}(\bar t,\bar t^\prime)c^{}_-(\bar t^\prime)\Biggr\}\Biggr ],
\nonumber
\end{eqnarray}
where the fermionic operators with the $+$ or $-$ subscript live on the corresponding time branch; the minus sign enters from the change of direction in how the time evolves on the two different branches.  When both time arguments lie on the $+$ branch or the $-$ branch, we have time-ordered ($T$) or anti-time-ordered $(\bar T)$ objects, respectively.  When one is on the $+$ and one on the $-$, we have have the lesser ($<$) or greater $(>)$ functions.  To be concrete, the four different Green's functions are defined as follows (we suppress the trace over the $f$-electrons, which is needed for the Falicov-Kimball model, but is neglected for the moment):
\begin{eqnarray}
 G^T(t,t^\prime)&=&-i{\rm Tr}_{c} \{\mathcal{T}_t \rho_{imp}S(\lambda_C)c_+^{}(t)c^\dagger_+(t^\prime)\}/\mathcal{Z},\\
G^{\bar T}(t,t^\prime)&=&-i{\rm Tr}_{c} \{\mathcal{T}_{\bar t} \rho_{imp}S(\lambda_C)c_-^{}(t)c^\dagger_-(t^\prime)\}/\mathcal{Z},\\
G^<(t,t^\prime)&=&i{\rm Tr}_{c} \{\rho_{imp}S(\lambda_C)c^\dagger_+(t^\prime)c_-^{}(t)\}/\mathcal{Z},\\
G^>(t,t^\prime)&=&-i{\rm Tr}_{c} \{\rho_{imp}S(\lambda_C)c_+^{}(t)c^\dagger_-(t^\prime)\}/\mathcal{Z},
\end{eqnarray}
where the $t$ and $\bar t$ subscripts denote time-ordering or anti-time-ordering, respectively.
The retarded and advanced Green's functions are defined to be $G^r=G^T-G^<$ and $G^a=-G^{\bar T}+G^<$, which can be written as
\begin{eqnarray}
 G^r(t,t^\prime)&=&-i\theta(t-t^\prime){\rm Tr}_{c}\left [ \rho_{imp}S(\lambda_C)\{c^{}(t),c^\dagger(t^\prime)\}_+\right ] \nonumber\\
\\
G^a(t,t^\prime)&=&i\theta(t^\prime-t){\rm Tr}_{c}\left [ \rho_{imp}S(\lambda_C)\{c^{}(t),c^\dagger(t^\prime)\}_+\right ] \nonumber\\
\end{eqnarray}
in terms of the operators.  Similarly, the Keldysh and anti-Keldysh Green's functions are defined to be $G^K=G^>+G^<$ and $G^{\bar K}=-G^T-G^{\bar T}+G^>+G^<$, or
\begin{eqnarray}
G^K(t,t^\prime)&=&-i{\rm Tr}_c \left \{ \rho_{imp}S(\lambda_C)[c^{}(t),c^{\dagger}(t^\prime)]_- \right \},\\
G^{\bar K}(t,t^\prime)&=&-i{\rm Tr}_c \left \{ \rho_{imp}S(\lambda_C)\{[c^{}(t),c^{\dagger}(t^\prime)]_-\right .\nonumber\\
&~&\left . -[c^{}(t),c^{\dagger}(t^\prime)]_-\}\right \}.
\end{eqnarray}
Note that the anti-Keldysh Green's function vanishes, but we need its functional form in order to take functional derivatives for different Green's functions below.
Using the definition of the evolution operator, we immediately find that the Green's functions can be found as functional derivatives of the partition function.  The explicit relations are
\begin{eqnarray}
 G^T(t,t^\prime)&=&-\frac{\delta \ln \mathcal{Z}}{\delta \lambda^T(t^\prime,t)},~~
G^{\bar T}(t,t^\prime)=-\frac{\delta \ln \mathcal{Z}}{\delta \lambda^{\bar T}(t^\prime,t)},\\
G^<(t,t^\prime)&=&\frac{\delta \ln \mathcal{Z}}{\delta \lambda^{>}(t^\prime,t)},~~~~
G^>(t,t^\prime)=\frac{\delta \ln \mathcal{Z}}{\delta \lambda^{<}(t^\prime,t)},
\end{eqnarray}
for the time-ordered, anti-time-ordered, lesser, and greater Green's functions, respectively.  We will discuss the alternative retarded, advanced, Keldysh and anti-Keldysh basis below.
Now that we have all of these definitions, we can actually solve explicitly for these Green's functions using the equations of motion (EOMs).  We take the defintion of the Green's function, and differentiate with respect to time; where appropriate, one needs to take into account the time ordering, which brings down terms proportional to the $\lambda$ fields.  This procedure is straightforward to complete, and the end result is the following:
\begin{eqnarray}
 i\partial_t G^T(t,t^\prime)&=&\delta(t-t^\prime)-\mu G^T(t,t^\prime)\nonumber\\
&+&\int_{-\infty}^\infty d\bar t \lambda^T(t,\bar t)G^T(\bar t,t^\prime)\nonumber\\
&-&\int_{-\infty}^\infty d\bar t \lambda^<(t,\bar t)G^>(\bar t,t^\prime),
\end{eqnarray}
\begin{eqnarray}
-i\partial_t G^{\bar T}(t,t^\prime)&=&\delta(t-t^\prime)+\mu G^{\bar T}(t,t^\prime)\nonumber\\
&-&\int_{-\infty}^\infty d\bar t \lambda^>(t,\bar t)G^<(\bar t,t^\prime)\nonumber\\
&+&\int_{-\infty}^\infty d\bar t \lambda^{\bar T}(t,\bar t)G^{\bar T}(\bar t,t^\prime),
\end{eqnarray}
\begin{eqnarray}
i\partial_t G^{<}(t,t^\prime)&=&-\mu G^{<}(t,t^\prime)
+\int_{-\infty}^\infty d\bar t \lambda^T(t,\bar t)G^<(\bar t,t^\prime)\nonumber\\
&-&\int_{-\infty}^\infty d\bar t \lambda^{<}(t,\bar t)G^{\bar T}(\bar t,t^\prime),
\end{eqnarray}
\begin{eqnarray}
-i\partial_t G^{>}(t,t^\prime)&=&\mu G^{>}(t,t^\prime)
-\int_{-\infty}^\infty d\bar t \lambda^>(t,\bar t)G^T(\bar t,t^\prime)\nonumber\\
&+&\int_{-\infty}^\infty d\bar t \lambda^{\bar T}(t,\bar t)G^{>}(\bar t,t^\prime),
\end{eqnarray}
with similar equations for derivatives with respect to $t^\prime$ which we do not write down explicitly.

On the lattice, one can exactly solve for the analogous contour-ordered Green's functions in the case where $U=0$~\cite{noneq_nonint}.  One can see from the exact solution, that the noninteracting retarded Green's function becomes average time independent for long times after the field has been turned on.  We make the ansatz that the retarded self-energy is also independent of average time in the long-time limit (which is consistent with gauge-invariance arguments~\cite{jauho,freericks_book,aoki}, but is an independent ansatz).  Then, it is straightforward to show that the interacting retarded Green's function is independent of average time.  In general, we can only show that the average time dependence of the contour-ordered Green's function is periodic in average time with a period given by the Bloch period $2\pi/E$~\cite{freericks_hubbard_prl}.  This allows us to make a continuous Fourier transform with respect to the relative time (frequency dependence $\omega$) and a discrete Fourier transform with respect to the average time (Fourier components $n\omega_B=nE$) for all of the different components of the contour-ordered objects (time-ordered, anti-time-ordered, lesser, greater, and Keldysh); the retarded and advanced components depend only on the continuous frequency $\omega$.

In order to find the Green's functions, we need to employ the simplification from the retarded and advanced Green's
functions, which depend only on the relative time, or frequency $\omega$.  Hence, we need the EOM for the retarded and advanced functions, which can easily be shown to satisfy
\begin{eqnarray}
 (i\partial_t+\mu)G^r(t)&=&\delta(t)+\int_{-\infty}^\infty d\bar t \lambda^r(t-\bar t)G^r(\bar t),\\
(i\partial_t+\mu)G^a(t)&=&\delta(t)+\int_{-\infty}^\infty d\bar t \lambda^a(t-\bar t)G^a(\bar t),
\end{eqnarray}
which are solved by 
\begin{equation}
 G^r(\omega)=\frac{1}{\omega+\mu-\lambda^r(\omega)},~
G^a(\omega)=\frac{1}{\omega+\mu-\lambda^a(\omega)};
\end{equation}
where we have solved the problem in frequency space after a Fourier transformation.
Next, we can show, by using the identities that relate the different quantities ($T$, $\bar T$, $<$, $>$, $r$, and $a$) to each other, that the EOM for $G^<$ becomes
\begin{eqnarray}
 (i\partial_t+\mu)G^<(t,t^\prime)&=&\int_{-\infty}^\infty d\bar t \lambda^r(t-\bar t)G^<(\bar t,t^\prime)\nonumber\\
&+&\int_{-\infty}^\infty d\bar t \lambda^<(t,\bar t)G^a(\bar t-t^\prime),
\end{eqnarray}
since $\lambda^T=\lambda^r+\lambda^<$ and $G^{\bar T}=G^<-G^a$. Substituting in the appropriate Fourier expansions then yields
\begin{eqnarray}
 &~&(\omega+\frac{n}{2}\omega_B)G^<(n\omega_B,\omega)=\lambda^r(\omega+\frac{n}{2}\omega_B)G^<(n\omega_B,\omega)\nonumber\\
&~&~~~~~~~+\lambda^<(n\omega_B,\omega)G^a(\omega-\frac{n}{2}\omega_B).
\end{eqnarray}
This can be directly solved to yield
\begin{equation}
 G^<(n\omega_B,\omega)=\frac{\lambda^<(n\omega_B,\omega)}{D(\omega+\frac{n}{2}\omega_B,\omega-\frac{n}{2}\omega_B)},
\end{equation}
with
\begin{eqnarray}
 &~&D(\omega+\frac{n}{2}\omega_B,\omega-\frac{n}{2}\omega_B)=[\omega+\frac{n}{2}\omega_B+\mu-\lambda^r(\omega+
\frac{n}{2}\omega_B)]\nonumber\\
&~&~~\times[\omega-\frac{n}{2}\omega_B+\mu-\lambda^a(\omega-\frac{n}{2}\omega_B)].
\end{eqnarray}
Similarly, we have
\begin{equation}
 G^>(n\omega_B,\omega)=\frac{\lambda^>(n\omega_B,\omega)}{D(\omega+\frac{n}{2}\omega_B,\omega-\frac{n}{2}\omega_B)},
\end{equation}
and $G^T(n\omega_B,\omega)=G^<(n\omega_B,\omega)+\delta_{n0}G^r(\omega)$ and $G^{\bar T}(n\omega_B,\omega)=G^<(n\omega_B,\omega)-\delta_{n0}G^a(\omega)$; the Kronnecker delta functions enter because the retarded and advanced Green's functions are independent of average time.
One can directly show that 
\begin{equation}
 \lambda^T(n\omega_b,\omega)=\lambda^{\bar T}(n\omega_b,\omega)=\lambda^<(n\omega_b,\omega)=\lambda^>(n\omega_b,\omega),
\end{equation}
for $n\ne 0$.
Now replacing the retarded and advanced quantities in the denominator by the time-ordered, anti-time-ordered, lesser
and greater quantities, gives
\begin{eqnarray}
 D&=&[\omega+\frac{n}{2}\omega_B+\mu-\lambda^T(0,\omega+\frac{n}{2}\omega_B)]\\
&\times&[\omega-\frac{n}{2}\omega_B+\mu-\lambda^{\bar T}(0,\omega-\frac{n}{2}\omega_B)]\nonumber\\
&+&\lambda^<(0,\omega+\frac{n}{2}\omega_B)[-n\omega_B+\lambda^>(0,\omega-\frac{n}{2}\omega_B)],\nonumber
\end{eqnarray}
which involves just the zeroth component of the Fourier series terms for the dynamical mean fields.

Finally, we can now solve the functional differential equations, which take the form $G(n\omega_b.\omega)=\pm\delta \ln \mathcal{Z}/\delta \lambda(-n\omega_B,\omega)$, and determine the partition function as
\begin{eqnarray}
 &~&\ln \mathcal{Z}=\int d\omega \ln \left [ (\omega+\mu-\lambda^T(0,\omega))(\omega+\mu\right .\nonumber\\
&~&~~~~+\left .\lambda^{\bar T}(0,\omega))+\lambda^<(0,\omega)\lambda^>(0,\omega)\right ]+C\\
&+&\sum_{n\ne 0}\frac{\lambda^<(n\omega_B,\omega)\lambda^>(-n\omega_B,\omega)-\lambda^T(n\omega_B,\omega)\lambda^{\bar T}(-n\omega_B,\omega)}{D(\omega+\frac{n}{2}\omega_B,\omega-\frac{n}{2}\omega_B)},\nonumber
\end{eqnarray}
where we fix an undetermined constant $C$ by equating with the noninteracting result.  It is a straightforward exercise 
to show that the functional derivatives of this partition function yield the Green's functions.  

Our next step is to repeat these results for the alternative $r$, $a$, $K$, $\bar K$ basis.  To start, the
evolution operator becomes
\begin{eqnarray}
 S(\lambda_C)&=&\mathcal{T}_t\exp\Biggr\{-\frac{i}{2}\int_{-\infty}^\infty dt \int_{-\infty}^\infty dt^\prime\nonumber\\
&~&[c_+^\dagger(t)-c_-^\dagger(t)]\lambda^r(t,t^\prime)[c_+(t^\prime)+c_-(t^\prime)]\nonumber\\
&+&[c_+^\dagger(t)+c_-^\dagger(t)]\lambda^a(t,t^\prime)[c_+(t^\prime)-c_-(t^\prime)]\nonumber\\
&+&[c_+^\dagger(t)-c_-^\dagger(t)]\lambda^K(t,t^\prime)[c_+(t^\prime)-c_-(t^\prime)]\nonumber\\
&+&[c_+^\dagger(t)+c_-^\dagger(t)]\lambda^{\bar K}(t,t^\prime)[c_+(t^\prime)+c_-(t^\prime)]\Biggr\},
\end{eqnarray}
where we have added a new field $\lambda^{\bar K}$ which will be set equal to zero for all physical matrix elements that we evaluate.  Using the definitions for the retarded, advanced, Keldysh and anti-Keldysh Green's functions shows that
we can determine them via functional derivatives
\begin{eqnarray}
 G^r(t,t^\prime)&=&-\frac{\delta \ln \mathcal{Z}}{\delta \lambda^r(t^\prime,t)},~~
G^a(t,t^\prime)=-\frac{\delta \ln \mathcal{Z}}{\delta \lambda^a(t^\prime,t)},\\
G^K(t,t^\prime)&=&-\frac{\delta \ln \mathcal{Z}}{\delta \lambda^{\bar K}(t^\prime,t)},~~
G^{\bar K}(t,t^\prime)=-\frac{\delta \ln \mathcal{Z}}{\delta \lambda^{K}(t^\prime,t)},
\end{eqnarray}
where all derivatives must be evaluated with $\lambda^{\bar K}=0$ after taking the derivatives. Since we have already determined the retarded and advanced Green's functions when we solved for the Green's functions in the other basis, we need only solve for the Keldysh and anti-Keldysh Green's functions, which can be easily determined (via the EOM, or using the relation between the lesser and greater Green's functions) to give
\begin{eqnarray}
&~& ~~G^K(n\omega_B,\omega)=\frac{\lambda^K(n\omega_B,\omega)}{D(\omega+\frac{n}{2}\omega_B,\omega-\frac{n}{2}\omega_B)},\\
 &~&~~G^{\bar K}(n\omega_B,\omega)=\frac{\lambda^{\bar K}(n\omega_B,\omega)}{D(\omega+\frac{n}{2}\omega_B,\omega-\frac{n}{2}\omega_B))}=0.
\end{eqnarray}
Using these solutions, we can integrate them to find the partition function, but we need to introduce appropriate anti-Keldysh fields to give us the final functional form, because we set that field to zero in all functional derivatives we evaluate.  The end result is
\begin{eqnarray}
\ln \mathcal{Z}&=&\int d\omega \ln \left [ (\omega+\mu-\lambda^r(\omega))(\omega+\mu-\lambda^{a}(\omega))\right .\nonumber\\
&-&\left .\lambda^K(0,\omega)\lambda^{\bar K}(0,\omega)\right ]+C\nonumber\\
&-&\sum_{n\ne 0}\frac{\lambda^K(n\omega_B,\omega)\lambda^{\bar K}(-n\omega_B,\omega)}{D(\omega+\frac{n}{2}\omega_B,\omega-\frac{n}{2}\omega_B)}.
\end{eqnarray}
This completes the derivations for the two equivalent bases for the impurity model with no localized electrons.

We now generalize for the case of the Falicov-Kimball model, where we include the trace over the localized particles in all of the relevant expectation values.  The changes are straightforward to work out, and we report them only in the $r$, $a$, $K$, and $\bar K$ basis. Define the effective partition function by (noting that $\lambda^r=\lambda^{a*}$)
\begin{equation}
\mathcal{Z}_0(\mu)=C^\prime\exp\left [ \int d\omega \frac{|\omega+\mu-\lambda^r(\omega)|^2}{|\omega+\mu|^2}\right ],
\end{equation}
where $C^\prime={\rm Tr}_c\rho_{imp}$, is the noninteracting partition function for the steady state in the absence of the time-dependent fields. Then the full partition function is
\begin{equation}
 \mathcal{Z}_{FK}=\mathcal{Z}_0(\mu)+e^{-\beta E_f}\mathcal{Z}_0(\mu-U),
\end{equation}
with $E_f$ the localized electron Fermi level, adjusted to give the correct average filling of the localized particles. The average filling satisfies
\begin{equation}
 w_1=1-w_0=\frac{e^{-\beta E_f}\mathcal{Z}_0(\mu-U)}{\mathcal{Z}_{FK}};~~w_0=\frac{\mathcal{Z}_0(\mu)}{\mathcal{Z}_{FK}}.
\end{equation}
The Green's functions take the same form as they had for the case with no localized particles, except we now have the sum of two terms: one, weighted by $w_0$, which is the result we have found above, and one, weighted by $w_1$, which has the same form as the result we derived above but with $\mu\rightarrow\mu-U$. The retarded Green's function takes the same functional form as it has in equilibrium, namely
\begin{equation}
 G^r(\omega)=\frac{1-w_1}{\omega+\mu-\lambda^r(\omega)}+\frac{w_1}{\omega+\mu-U-\lambda^r(\omega)}.
\end{equation}
The Keldysh Green's function is more complicated:
\begin{eqnarray}
 G^K(n\omega_B,\omega)&=&\lambda^K(n\omega_B,\omega)\Biggr [ \frac{1-w_1}{D(\omega+\frac{n}{2}\omega_B,\omega-\frac{n}{2}\omega_B)}\nonumber\\
&+&\frac{w_1}{D(\omega+\frac{n}{2}\omega_B,\omega-\frac{n}{2}\omega_B)|_{\mu\rightarrow\mu-U}}\Biggr ].
\end{eqnarray}

Note that the retarded Green's function appears to be independent of the Keldysh Green's function, and hence has
no explicit dependence on the distribution function, and thereby, no dependence on the steady-state density matrix,
but one needs to carefully check to see whether the functional derivative of $w_1$ has any dependence on $\lambda^K$.  Indeed, it is very easy to show that there is no such dependence, because any functional derivative of the partition
function with respect to $\lambda^K$ will bring down a factor of $\lambda^{\bar K}$, which is set equal to zero, so we immediately learn that
\begin{equation}
 \left .\frac{\delta^m G^r(\omega)}{\delta \lambda^K(n_1\omega_B,\omega^\prime_1)\ldots
\delta \lambda^K(n_m\omega_B,\omega^\prime_m)}\right |_{\lambda^{\bar K}=0}=0,
\end{equation}
for all powers $m$.  {\it This is the generalization of the proof that the conduction-electron density of states for the Falicov-Kimball model is independent of temperature in equilibrium to a proof of the distribution-function independence of the conduction-electron density of states in the nonequilibrium ``steady state''.}  Of course, this proof only holds in the case of a canonical distribution of the heavy particles, where their filling is kept constant as the temperature or the nonequilibrium driving fields are varied.

In order to understand this result more fully, we examine explicitly the functional derivative of the retarded Green's function with respect to the Keldysh dynamical mean field in the time basis.  We find that
\begin{eqnarray}
 &~&\frac{\delta G^r(t,t^\prime)}{\delta \lambda^K(\bar t^\prime,\bar t)}=\frac{1}{4}{\rm Tr}_{cf}\mathcal{T}_C\rho_{imp}S(\lambda_C)\nonumber\\
&~&~~\times[c^\dagger_+(t^\prime)-c^\dagger_-(t^\prime)][c^{}_+(t)+c^{}_-(t)]\nonumber\\
&~&~~\times
[c^\dagger_+(\bar t^\prime)-c^\dagger_-(\bar t^\prime)][c^{}_+(\bar t)-c^{}_-(\bar t)]/\mathcal{Z}_{FK}.
\end{eqnarray}
The $+$ and $-$ subscripts denote which branch of the Keldysh contour the time arguments lie on.  This expression can be simplified somewhat to
the form
\begin{eqnarray}
 &~&\frac{\delta G^r(t,t^\prime)}{\delta \lambda^K(\bar t^\prime,\bar t)}=\frac{1}{4}{\rm Tr}_{cf}\rho_{imp}S(\lambda_C)\label{eq: fcnl_deriv}\\
&~&\times\Big \{ -2c(t)\mathcal{T}_t[c^\dagger(t^\prime)c^\dagger(\bar t^\prime)c(\bar t)]+2\mathcal{T}_{\bar t}[c^\dagger(t^\prime)c^\dagger(\bar t^\prime)c(\bar t)]c(t)\nonumber\\
&~&+\mathcal{T}_t[c^\dagger(t^\prime)c(t)]\mathcal{T}_{\bar t}[c^\dagger(\bar t^\prime)c(\bar t)]+
\mathcal{T}_t[c^\dagger(t^\prime)c^\dagger(\bar t^\prime)]\mathcal{T}_{\bar t}[c(t)c(\bar t)]\nonumber\\
&~&+\mathcal{T}_t[c(t)c^\dagger(\bar t^\prime)]\mathcal{T}_{\bar t}[c^\dagger(t^\prime)c(\bar t)]-
\mathcal{T}_t[c^\dagger(t^\prime)c(\bar t)]\mathcal{T}_{\bar t}[c(t)c^\dagger(\bar t^\prime)]\nonumber\\
&~&-\mathcal{T}_t[c(t)c(\bar t)]\mathcal{T}_{\bar t}[c^\dagger(t^\prime)c^\dagger(\bar t^\prime)]-
\mathcal{T}_t[c^\dagger(\bar t^\prime)c(\bar t)]\mathcal{T}_{\bar t}[c^\dagger(t^\prime)c(t)]\Big \}.\nonumber
\end{eqnarray}
We know that this correlation function vanishes for the Falicov-Kimball model, but it is nontrivial to show that this is so, and it does not appear obvious at all from the operator expectation value above.

Note that we cannot make any further progress on evaluating the Keldysh Green's function in the steady state without knowing what the density matrix is, and we do not know this explicitly.

\section{Formalism for the Hubbard model}

In the case of the Hubbard model, both particles can move, so we have an evolution operator given by the product $S(\lambda_C^\uparrow)S(\lambda_C^\downarrow)$ in all of the expectation values that we need to evaluate; the $\uparrow$ electrons are the old $c$ electrons and the $\downarrow$ electrons are the old $f$ electrons. Now we can set up the formalism for calculating the Green's functions directly, just as for the Falicov-Kimball model, but unfortunately, in this case, the equations of motion cannot be solved explicitly, so we cannot determine an explicit formula for the partition function anymore.  These expectation values can be calculated with the numerical renormalization group, if we have the explicit operator form for the density matrix, but as we discussed above, this is not the case.

What we do have is an explicit operator average that tells us the dependence of the Green's function on the distribution function (via the Keldysh dynamical mean field).  It is just the result in Eq.~(\ref{eq: fcnl_deriv}), generalized to include spin indices $\sigma$ on the operators with unbarred time arguments, and $\sigma^\prime$ on operators with barred time arguments.  If that correlation function could be evaluated for an (approximate) solution of the retarded Green's function for the Hubbard model~\cite{freericks_hubbard_prl}, then we could determine how strong the distribution function dependence of the retarded Green's function was for that model.  Unfortunately, this does not appear to be a simple task at this point in time.

\section{Conclusions}

In this work, we have shown a number of explicit details for the impurity problem in nonequilibrium dynamical mean-field theory when one approaches the steady-state response.  For the Falicov-Kimball model, we can advance the functional analysis to the point where we can prove that the retarded (and advanced) Green's function does not depend on the distribution function.  In addition, we derived a two-particle correlation function, that would need to be evaluated for the general case, which determines the strength of the distribution-function dependence of the retarded (and advanced) Green's function.  While we can explicitly show this correlation function vanishes for the Falicov-Kimball model, we do not know how one would evaluate it for the Hubbard model, particularly because we do not know what the steady-state density matrix is for the model. Nevertheless, we anticipate that it is not large, as we expect the steady-state
nonequilibrium density of states to depend only weakly on the distribution function.

\section{Acknowledgments}

We acknowledge support from the National Science Foundation under grant number DMR-0705266.  I would like to also recognize useful discussions with F.~Anders, H.~Krishnamurthy, A.~Joura, and V.~Turkowski.

\end{document}